\documentclass[journal=jacsat,manuscript=article]{achemso}

\usepackage[version=3]{mhchem} 
\usepackage{bm} 

\author{Yuuki Ishiwatari}
\email{ishiwatari-527.uki@keio.jp}
\affiliation[Keio UniversityM]
{Department of Mechanical Engineering, Keio University, Yokohama}
\author{Takahiro Yokoyama}
\affiliation[Keio UniversityM]
{Department of Mechanical Engineering, Keio University, Yokohama}
\author{Tomoya Kojima}
\affiliation[Keio UniversityC]
{Department of Applied Chemistry, Keio University, Yokohama}
\author{Taisuke Banno}
\affiliation[Keio UniversityC]
{Department of Applied Chemistry, Keio University, Yokohama}
\author{Noriyoshi Arai}
\email{arai@mech.keio.ac.jp}
\phone{+81 (45)566 1846}
\affiliation[Keio UniversityM]
{Department of Mechanical Engineering, Keio University, Yokohama}

\title[title]
  {Composition-agnostic prediction of self-assembly in multicomponent amphiphile mixtures from molecular structure}

\abbreviations{CPP, DPD, ML, NN, CNN, GRU, RNN, GCN, MLP}

\begin{document}






\begin{abstract}
Predicting self-assembly in multi-component amphiphilic systems is challenging due to the complexity of intercomponent interactions and the combinatorial growth of possible formulations. In this study, we develop a unified machine-learning framework that directly predicts self-assembly behavior from the molecular structures of constituent components, independent of the number or identity of those components. We extend the critical packing parameter (CPP) to multi-component systems and generate a large dataset of self-assembled morphologies using dissipative particle dynamics (DPD) simulations. By systematically evaluating twelve combinations of feature extraction methods and model architectures, we find that models incorporating a fully connected graph convolutional network (GCN) layer achieve superior performance, with the GCN–GCN architecture accurately capturing both intramolecular relationships and intercomponent interactions. Notably, this model exhibits strong extrapolative capability: it accurately predicts CPP values for five-component mixtures even when trained only on systems with fewer components, and it maintains high accuracy for mixtures composed of molecular species that are entirely absent from the training data. These results demonstrate that a composition-agnostic predictive framework can enable efficient virtual screening and provide a foundation for the rational design of complex amphiphilic materials.
\end{abstract}


\section{Introduction}
Self-assembly of amphiphilic molecules underlies a broad range of functional soft materials, from drug-delivery carriers\cite{Pranav2023,Gao2020} and gene-delivery vectors\cite{Zhao2014,Wang2024-2} to detergents, \cite{Kapitanov2019,Perinelli2020} emulsifiers, \cite{Kim2024,Zhao2023} and stimuli-responsive amphiphilic systems.\cite{Jacoby2021,Yu2014}
The resulting morphologies can undergo dramatic transitions in response to subtle variations in molecular structure or environmental conditions, often leading to orders-of-magnitude changes in macroscopic properties.
For example, a temperature-induced transition from worm-like micelles to multi-lamellar vesicles has been reported to decrease the complex viscosity by more than two orders of magnitude. \cite{Wang2012}
Such observations highlight the intimate connection between self-assembled morphology and macroscopic function, emphasizing that predictive control over self-organization is essential for rational soft-matter design.

In recent years, functional materials have increasingly relied on multi-component formulations.
Mixing chemically distinct amphiphiles enables access to richer self-assembled structures and advanced functionalities that cannot be achieved by single-component systems.
Examples include lipid nanoparticle systems,\cite{Hassett2019, Akinc2010} polymer–surfactant complexes,\cite{Zhu2021, AbbasiMoud2022} and hybrid materials involving colloids or nanoparticles.\cite{Laghrissi2024, Heo2024}
However, multi-component design also introduces major challenges.
As the number of components increases, intercomponent interactions become more complex and nonlinear, and the combinatorial design space grows exponentially.
Experimental exploration of such high-dimensional formulation spaces is extremely costly, making predictive design of self-assembly and the resulting material functions substantially more difficult.
Current multi-component material development therefore depends heavily on empirical trial-and-error, which slows discovery and generates substantial waste, increasing the demand for data-driven molecular and formulation design strategies.

Against this backdrop, machine learning (ML) has emerged as a powerful tool for accelerating molecular design in chemistry and materials science.
ML models have demonstrated strong predictive performance for a wide range of molecular properties, including acid dissociation constants,\cite{Roszak2019,An2024,Dobbelaere2024} interaction energies,\cite{Gelžinytė2024,Pereira2017,Choi2022} and biological activities,\cite{Ahmadi2024,Li2023,Nguyen-Vo2020} and their application to multi-component chemical systems\cite{Coley2019,Sarlar2021,Qiao2021} is steadily increasing.
For instance, Hanaoka accurately predicted glass-transition temperatures, boiling points, and densities of copolymers and binary or ternary mixtures.\cite{Hanaoka2020}
Similarly, Qin et al. successfully predicted composition-dependent activity coefficients with accuracy exceeding that of widely used thermodynamic models such as UNIFAC.\cite{Qin2023}
Despite these advances, existing ML approaches for multi-component systems typically require separate models for each number of components or composition type, limiting generality. 
Critically, no current framework is capable of treating systems with varying numbers of components in a unified manner or extrapolating to mixtures containing molecules absent from the training set—capabilities that are indispensable for broadly applicable materials design.

To address these challenges, we develop a unified ML model that predicts the self-assembled morphology of multi-component mixtures directly from molecular structures.
Our framework extends our previously established single-component prediction method based on the critical packing parameter (CPP)\cite{Ishiwatari2024} to mixtures, enabling morphology prediction using only molecular representations as inputs.
A key requirement for such a model is its ability to capture the hierarchical nature of self-assembly, from intramolecular structure to intercomponent interactions and emergent mesoscale organization. We therefore systematically evaluate combinations of machine learning algorithms and architectural designs to identify the representational and structural features necessary for composition-agnostic prediction.
The resulting model operates independently of the number of components and reliably extrapolates to unknown molecular combinations, enabling a general predictive framework for multi-component self-assembly.
This capability opens the door to high-throughput virtual screening, reduced reliance on iterative experimental formulation, and more sustainable pathways for the discovery and optimization of soft materials.

\section{Methods}

\subsection{Molecular simulation}
We employed the dissipative particle dynamics (DPD) method,\cite{Hoogerbrugge1992,Espanol1995,Groot1997,Santo2021} a coarse‑grained simulation technique.
To build efficient machine learning models for the prediction of complex self-assembled structures, it is necessary to prepare a large and diverse set of structures as training data.
Generating such structures requires a simulation method that can efficiently capture mesoscale aggregation behavior while remaining computationally tractable for high-throughput sampling. 
DPD satisfies these requirements: it can reproduce the essential hydrodynamic and thermodynamic features of self-assembly at a fraction of the cost of atomistic molecular dynamics. For this reason, DPD was selected as the simulation engine to construct the multi-component data set used in this work.

In DPD, the dynamics of each bead are governed by Newton’s equation of motion, incorporating conservative, dissipative, and random pairwise forces. The equation of motion for particle \(i\) is expressed as:

\begin{equation}
	m \frac{\text{d} {\textbf{v}_i}}{\text{d}t} = {\textbf{f}_i} = \sum_{j \neq i} {\textbf{F}}_{ij}^\mathrm{C} + \sum_{j \neq i} {\textbf{F}}_{ij}^\mathrm{D} + \sum_{j \neq i} {\textbf{F}}_{ij}^\mathrm{R}\;\;
  \label{eq:eq_motion}
  \end{equation}

Here, \(m\) denotes the bead mass, \(\mathbf{v}_i\) the velocity,  
\(\mathbf{F}_{ij}^{\mathrm{C}}\) the conservative force,  
\(\mathbf{F}_{ij}^{\mathrm{D}}\) the dissipative force, and  
\(\mathbf{F}_{ij}^{\mathrm{R}}\) the pairwise random force.
The conservative force \(\mathbf{F}_{ij}^{\mathrm{C}}\) is defined as:

 \begin{equation}
	{\textbf{F}}_{ij}^\mathrm{C} =
		\begin{cases}
			a_{ij} \left( 1-\dfrac{ \left| \textbf{r}_{ij}\right|}{r_{\mathrm c}} \right) \textbf{n}_{ij}, & \left| \textbf{r}_{ij} \right| \leq r_{\mathrm c} \\
			                      \;\;\;\;\;\;\;\;\;\;\;\;\;\;\;0,	& \left| \textbf{r}_{ij} \right| > r_{\mathrm c}\;\;,
		\end{cases}
	\label{eq:FC}
\end{equation}

\noindent where $\textbf{r}_{ij} = \textbf{r}_{j} - \textbf{r}_{i}$ is the vector between particles $i$ and $j$, $\textbf{n}_{ij} = \textbf{r}_{ij} / \left| \textbf{r}_{ij} \right|$ is the unit vector, $a_{ij}$ controls the repulsive interaction between particles $i$ and $j$, and $r_{\mathrm c}$ is the cutoff radius for force interactions.
The dissipative force (${\textbf{F}}_{ij}^\mathrm{D}$) and the random force (${\textbf{F}}_{ij}^\mathrm{R}$) are respectively given as:

\begin{equation}
	\label{eq:FD}
	{\textbf{F}}_{ij}^\mathrm{D} =
		\begin{cases}
			- \gamma \omega^{\mathrm D} \left( \left| \textbf{r}_{ij} \right| \right) \left( \textbf{n}_{ij} \cdot \textbf{v}_{ij} \right) \textbf{n}_{ij}, & \left| \textbf{r}_{ij} \right| \leq r_{\mathrm c},  \\
			                      \;\;\;\;\;\;\;\;\;\;\;\;\;\;\;0,	& \left| \textbf{r}_{ij} \right| > r_{\mathrm c}
		\end{cases}
\end{equation}

\begin{equation}
	\label{eq:FR}
	{\textbf{F}}_{ij}^\mathrm{R} =
		\begin{cases}
			\sigma \omega^{\mathrm R} \left( \left| \textbf{r}_{ij}\right| \right) \zeta_{ij} \Delta t^{-1/2} \textbf{n}_{ij}, & \left| \textbf{r}_{ij} \right| \leq r_{\mathrm c}\\
			                      \;\;\;\;\;\;\;\;\;\;\;\;\;\;\;0,	& \left| \textbf{r}_{ij} \right| > r_{\mathrm c}
		\end{cases}
\end{equation}

\noindent where $\textbf{v}_{ij} = \textbf{v}_{j} - \textbf{v}_{i}$, \(\sigma\) and \(\gamma\) are the noise and friction parameters, and $\zeta_{ij}$ is a random number based on a Gaussian distribution.
The weighting functions $\omega^{\mathrm R}$ and $\omega^{\mathrm D}$ depend on the distance as follows:

 \begin{equation}
  \label{eq:w_func}
	\omega^{D} \left( r \right) = \left[ \omega^{R} \left( r \right) \right]^2 =
		\begin{cases}
			\left[1 - \dfrac{\left| \bm{r}_{ij} \right|}{r_c}\right]^2,	& r_{ij} \leq r_c \\
			                      \;\;\;\;0,	& r_{ij} > r_c \;\;
		\end{cases}
\end{equation}

Temperature is controlled through the balance of random and dissipative forces, constrained by the fluctuation–dissipation theorem:

 \begin{equation}
  \label{eq:fd_theory}
	\sigma ^2 = 2 \gamma k_\mathrm{B} T,
\end{equation}

\noindent where $k_{\mathrm B}$ is Boltzmann's constant and $T$ is the temperature.

Bonded interactions between connected beads within amphiphilic molecules are represented using a harmonic spring force,
\begin{equation}
	{\textbf{F}}_{ij}^\mathrm{S} =
			-k_{s} \left( 1-\dfrac{ \left| \textbf{r}_{ij}\right|}{r_{\mathrm s}} \right) \textbf{n}_{ij}
	\label{eq:FS}
\end{equation}

\noindent with equilibrium bond distance \(r_s = 0.86 r_c\) and spring constant \(k_s = 100 k_{\mathrm B}T/r_{\mathrm c}^2\).

DPD simulations typically use reduced units, where lengths are scaled by \(r_{\mathrm c}\), masses by \(m\), and energies by \(k_{\mathrm B}T\).
The corresponding time unit is  \(\tau = r_{\mathrm c}\sqrt{m/(k_{\mathrm B}T)}\).
To correlate the simulation to physical units, a scaling procedure for length and time units is applied.\cite{Groot2001} Coarse-graining three water molecules into a single DPD bead gives each particle an effective mass of 54 atomic mass units. The particle density in the simulation is set to $\rho = 3 r_{\mathrm c}^{-3}$, implying that three DPD beads occupy a cubic volume of $r_{\mathrm c}^3$. Given that the molecular volume of a single water molecule is $0.03 \text{nm}^3$, this mapping yields the length unit $r_{\mathrm c}$ is $\left( 0.27 \text{nm}^{3} \right)^{1/3} \approx 0.6463 \text{nm}$. Using this length scale, along with the bead mass and thermal energy at room temperature, the corresponding DPD time unit is estimated to be \(\tau \approx 88\ \mathrm{ps}\).

\subsection{Simulation models and conditions}
\begin{figure}[tb]
    \centering
    \includegraphics[width=8cm]{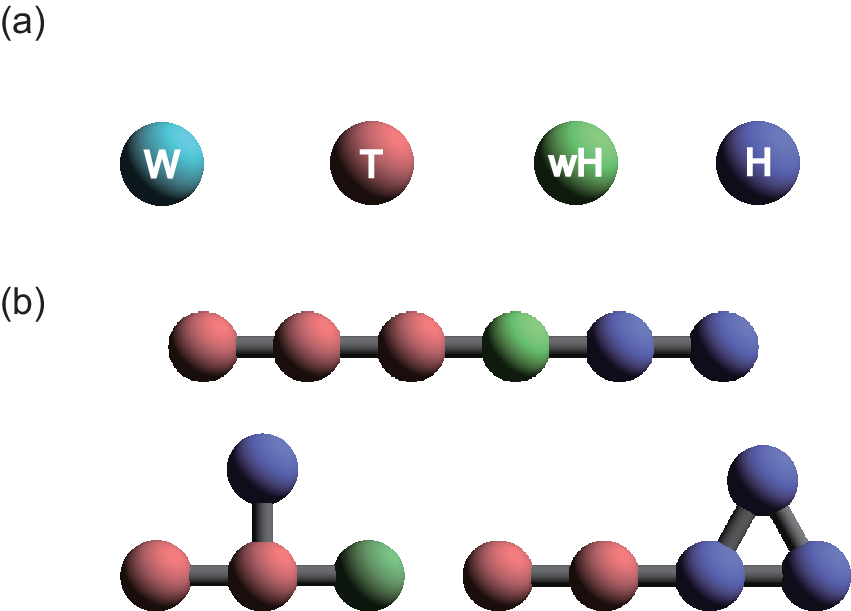}
    \caption{Particle model used in the simulations. (a) Coarse-grained DPD beads used in this study: water bead as solvent (cyan), hydrophobic tail bead (red), weakly hydrophilic head bead (green), and hydrophilic head bead (blue), labeled as W, T, wH, and H respectively. (b)Representative modeled amphiphilic molecules composed of three types of DPD beads: hydrophobic tail (red), weakly hydrophilic head (green), and hydrophilic head (blue). }
    \label{fig:model}
\end{figure}

The coarse-grained model used in this study is shown in Fig.\ref{fig:model}. A water molecule (W) is represented as a single coarse-grained particle as shown in Fig.\ref{fig:model}(a). In the amphiphilic molecular model, the red, green, and blue particles correspond to hydrophobic tail groups (T), weakly hydrophilic head groups (wH), and hydrophilic head groups (H), respectively. 
The type of wH particle was introduced to represent intermediate levels of hydrophilicity often found within amphiphilic molecules. Real surfactants frequently contain head groups with graded polarity, for example, partially hydrated functional groups or segments with weaker hydrogen-bonding ability.
Incorporating wH allows the model to capture such hydrophilicity differences in a simplified manner which leads to nonlinear variations in self-assembly behavior.
Typical amphiphilic molecular models are shown in Fig.\ref{fig:model}(b). A wide variety of models of amphiphilic molecules was created by changing the number of coarse-grained beads and the arrangement of particles with different hydrophobicities.
These models include linear, ring, and branched structures. In total, 479 amphiphilic molecular models were used in the study. These were combined with other molecular models to conduct simulations of multi-component systems containing \(n_\text{c} = 2-5\) non-water components, where \(n_\text{c}\) denotes the number of non-water components in the system.  
The number of simulated systems for each \(n_\text{c}\) is summarized in Table~\ref{tbl:dataset}.
\begin{table}[t]
\centering
\caption{Number of simulated systems for each number of non-water components \(n_\text{c}\).}
\label{tbl:dataset}
\begin{tabular}{cc}
\hline
\textbf{Number of components} & \textbf{Number of simulated systems} \\
\hline
\(n_\text{c} = 1\) & 497 \\
\(n_\text{c} = 2\) & 1,440 \\
\(n_\text{c} = 3\) & 1,440 \\
\(n_\text{c} = 4\) & 1,440 \\
\(n_\text{c} = 5\) & 480 \\
\hline
\end{tabular}
\end{table}

The non bonded interaction parameters ($a_{ij}$) between each pair of particles are shown in Table\ref{tbl:tbl1}. 
The concentration of the aqueous solution is $5~\mathrm{wt}\%$.
The total number of coarse-grained DPD beads in each simulation was fixed at 81,000.
Thus, approximately 4,050 beads corresponded to amphiphilic molecules and 76,950 to water beads; however, the exact numbers varied slightly depending on the number of beads composing each amphiphilic molecule (see SI for the details).
Initial configuration was prepared by randomly placing the water and amphiphilic molecules in the system. The simulation box has a volume of $30 \times 30 \times 30 r_{\mathrm c}^3$, and periodic boundary conditions were applied in all directions. All simulations were conducted in a constant-volume and constant-temperature ensemble until the equilibrium state was reached in $16,000\tau$.

\begin{table}[tb]
\small
  \caption{\ Interaction parameters $a_{ij}$ (in $k_{\mathrm B}T/r_{\mathrm c}$ unit) in DPD calculations}
  \label{tbl:tbl1}
  \begin{tabular*}{0.48\textwidth}{@{\extracolsep{\fill}}lllll}
    \hline
     & W & T & wH & H\\
    \hline
    W & 25.0 & 75.0 & 50.0 & 25.0\\
    T &      & 25.0 & 75.0 & 75.0\\
    wH &      &      & 25.0 & 50.0\\
     H &      &      &      & 25.0\\
    \hline
  \end{tabular*}
\end{table}

\subsection{CPP calculation (Target variable setup)}
The critical packing parameter (CPP) provides a quantitative link between the chemical and physical properties of amphiphiles and their resulting self-assembled structures.\cite{Israelachvili2011} Defined as a dimensionless measure, the CPP captures the geometric balance between hydrophilic and hydrophobic regions at the interface of self-assembled systems.
We calculated the CPP of the equilibrium self-assembled structures using our previously developed method.\cite{Ishiwatari2024}
The CPP is defined as 
\begin{equation}
    \label{eq:cpp}
    \text{CPP} = \frac{v}{a_0 l_c}.
\end{equation}
Here, $v$ was calculated by taking the volume of a single particle and multiplying it by the number of hydrophobic particles per molecule. 
We find the volume of a single particle assuming a uniform distribution at density $\rho=3 r_{\mathrm c}^{-3}$ with a fixed number of particles. 
The $a_0$ was obtained by multiplying the number of hydrophilic particles in contact with water by the surface area per particle. 
We determined the surface area per particle by raising the particle volume to the two-thirds power. 
The $l_c$ value was defined as the distance between the farthest hydrophobic particle and the closest hydrophilic particle, measured from the center of mass of the molecule. 

\begin{figure}[tb]
    \centering
    \includegraphics[width=8cm]{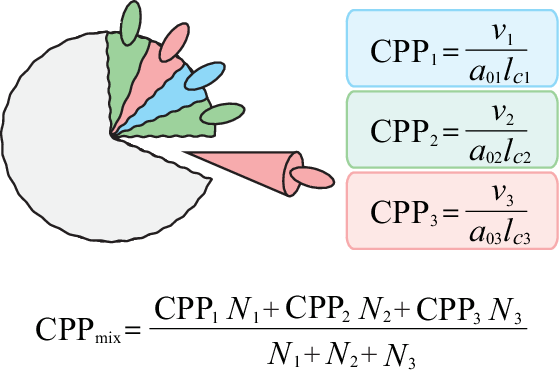}
    \caption{Calculation of CPP for multi-component systems: The CPP for each component is calculated and then averaged with weights given by the number of molecules $N_i$ in the cluster. An example for a 3-component system is shown.}
    \label{fig:cpp}
\end{figure}

Because the CPP was originally defined for single-solute systems, we modified it for multi-component systems as expressed in Eq.\ref{eq:cpp_mix}. 
\begin{equation}
    \label{eq:cpp_mix}
    \text{CPP}_{\mathrm{mix}} = 
    \frac{\displaystyle \sum_{i=1}^{n} \text{CPP}_{i} N_{i}}
         {\displaystyle \sum_{i=1}^{n} N_{i}}
\end{equation}
Figure \ref{fig:cpp} provides an illustrative example of the CPP calculation for a three-component system and demonstrates how the mixing rule is applied.
First, we calculate the CPP for each non-water component, then take a weighted average based on the number of molecules of each non-water component in the cluster. 

\begin{figure}[tb]
    \centering
    \includegraphics[width=18cm]{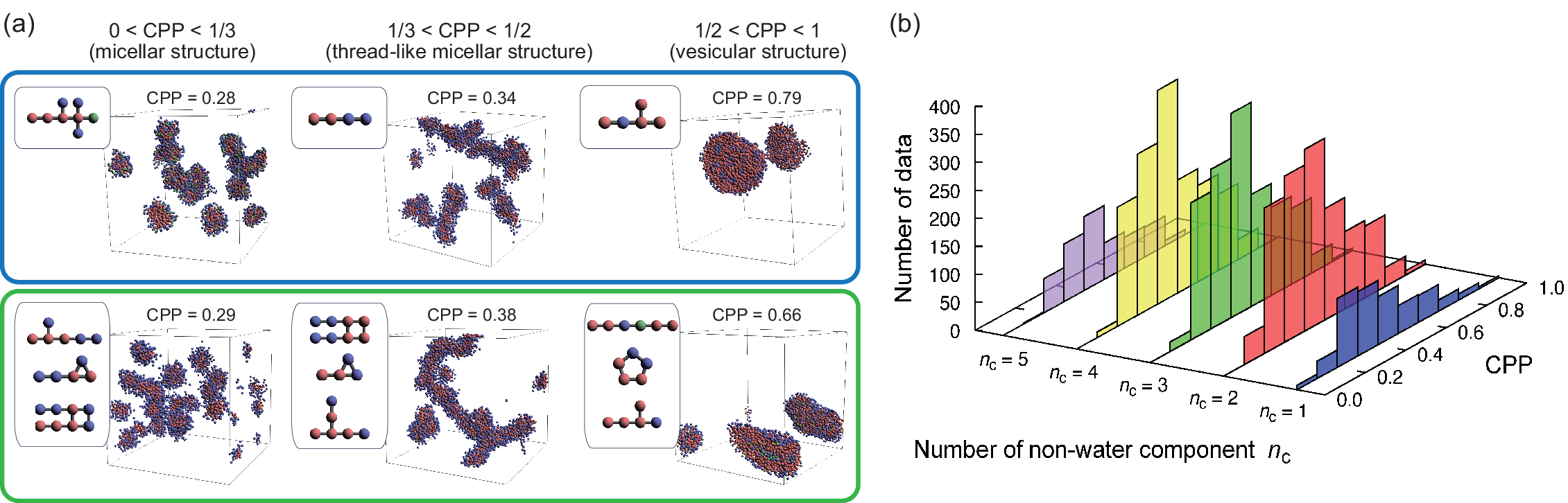}
    \caption{Summary of the calculated CPP values for the simulated self-assembled structures. 
    (a) Representative snapshots of self-assembled structures for systems with \(n_\text{c} = 1\) and \(n_\text{c} = 3\), together with the corresponding molecular models. These examples show that the computed CPP values are consistent with the expected morphologies.
    (b) Histograms of the CPP values for each component, binned at intervals of 0.1. Across systems with \(n_\text{c} = 1-5\), the CPP distributions cover the full range from 0 to 1, demonstrating the diversity of the CPP data generated for subsequent machine-learning analysis.}
    \label{fig:histo_snapshot}
\end{figure}

After computing the CPP values for multi-component systems using the procedure described above, we verified that the CPPs calculated from the obtained self-assembled structures fall within the ranges defined for their respective morphologies. As shown in Fig. \ref{fig:histo_snapshot}(a), micellar, thread-like micellar, and vesicular structures generated in the simulations yielded CPP values in the ranges \(0 < \mathrm{CPP} < 1/3\), \(1/3 < \mathrm{CPP} < 1/2\), and \(1/2 < \mathrm{CPP} < 1\), respectively, demonstrating that the computed CPPs are consistent with the established CPP-morphology definitions.

Furthermore, we calculated the mixed-system CPP for all self-assembled structures obtained in the simulations and summarized the results as histograms, as shown in Fig.~\ref{fig:histo_snapshot}(b). The CPP values were binned at intervals of 0.1. Across systems with \(n_\text{c} = 1-5\), the resulting CPP distribution spanned the full range from 0 to 1.
To construct the multi-component systems, the single-component dataset was first partitioned into three groups based on their CPP values (low, medium, and high), and multi-component mixtures were generated by randomly selecting components from these groups. Because this grouping-based sampling strategy does not frequently combine amphiphiles with extremely low or extremely high CPP values, the resulting CPP distribution contains relatively few samples near the lower and upper bounds of the CPP range.

\subsection{Molecular structure representation}
\begin{figure}[tb]
    \centering
    \includegraphics[width=8cm]{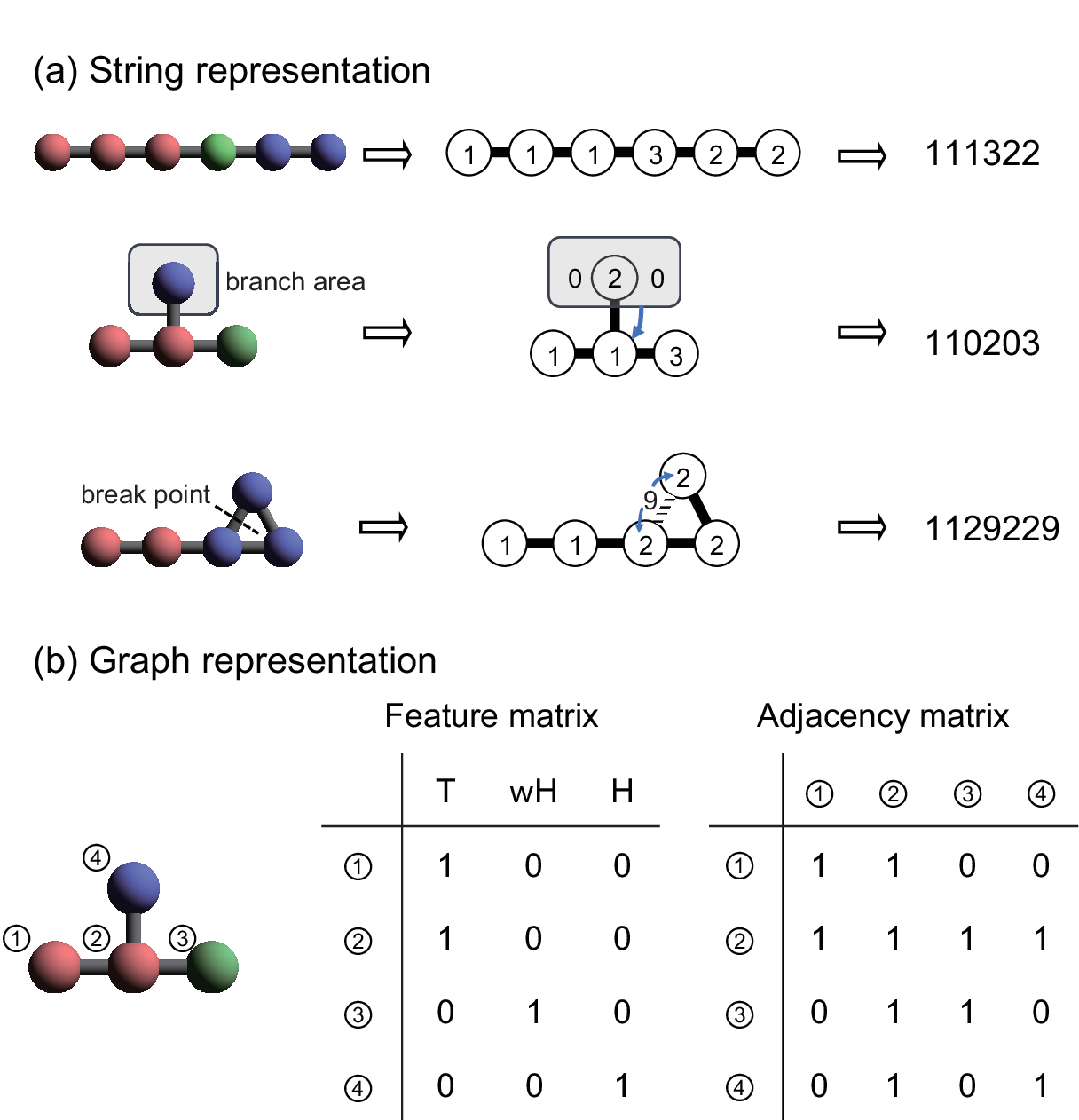}
    \caption{Two types of methods for encoding molecular structures used in this study: (a) Modified-SMILES, a method that linearly transforms coarse-grained molecular models, used as input for NN, CNN, and GRU algorithms; (b) Graph representation, a method that represents molecular structures as graph data using a feature matrix and an adjacency matrix, used as input for GCN algorithms.}
    \label{fig:encode}
\end{figure}

To make molecular structures readable by machine learning models, we used two encoding methods as shown in Fig.\ref{fig:encode}. One of them is a linear transformation method for molecular structures based on SMILES (Simplified Molecular Input Line Entry System), a widely used format for representing chemical structures as linear strings [Fig.\ref{fig:encode}(a)]. 
We developed this method in our previous work and named it “modified-SMILES”. To better suit coarse-grained molecular models, we simplified the conventional SMILES rules while retaining essential structural information, such as straight chains, branching, and rings. Further details can be found in our previous work.\cite{Ishiwatari2024}
The second encoding method represents the molecular structure as graph data. Graph data is a structure that describes relationships between objects using nodes and edges. In this study, particles are represented as nodes using a feature matrix, and bonds are represented as edges using an adjacency matrix [Fig.\ref{fig:encode}(b)].

\subsection{Machine learning model and architecture}
This study aims to predict the self-assembled structures of multi-component systems based solely on the molecular structures of their constituent components.  
To identify machine-learning approaches capable of extracting chemically meaningful features and capturing the hierarchical interactions that arise within and between components, we systematically explored multiple algorithms and architectural designs.
Specifically, four feature-extraction algorithms were examined; neural networks (NN),\cite{Rosenblatt1958,Rumelhart1986} convolutional neural networks (CNN),\cite{Fukushima1980,LeCun1989} gated recurrent units (GRU),\cite{Cho2014} a widely used variant of recurrent neural networks (RNN), and graph convolutional networks (GCN),\cite{Kipf2017} where each offers distinct representational capabilities for molecular data. 
NN, inspired by biological neural systems, is effective for capturing nonlinear relationships in data through layered transformations of weighted nodes.
CNN is designed to extract local spatial features through convolution and pooling operations. Although originally developed for image recognition, one-dimensional CNN can be applied to molecular string representations; in this study, they were used to learn local structural patterns within modified-SMILES sequences.
GRU, a gated RNN architecture, is well suited for sequential data such as time-series signals and chemical string representations. Our previous work demonstrated that GRU achieved high accuracy in predicting self-assembled structures in single-component systems. Here, modified-SMILES sequences were processed using a bidirectional GRU to capture both forward and backward dependencies.
GCN performs convolutions over graph topologies by aggregating features from each node and its neighbors. When molecules are represented as graphs with atoms as nodes and bonds as edges, GCN can naturally learn both local (bond-level) and global (molecule-level) structural information, making them particularly appropriate for chemical systems.

\begin{figure}[htb]
    \centering
    \includegraphics[width=0.8\linewidth]{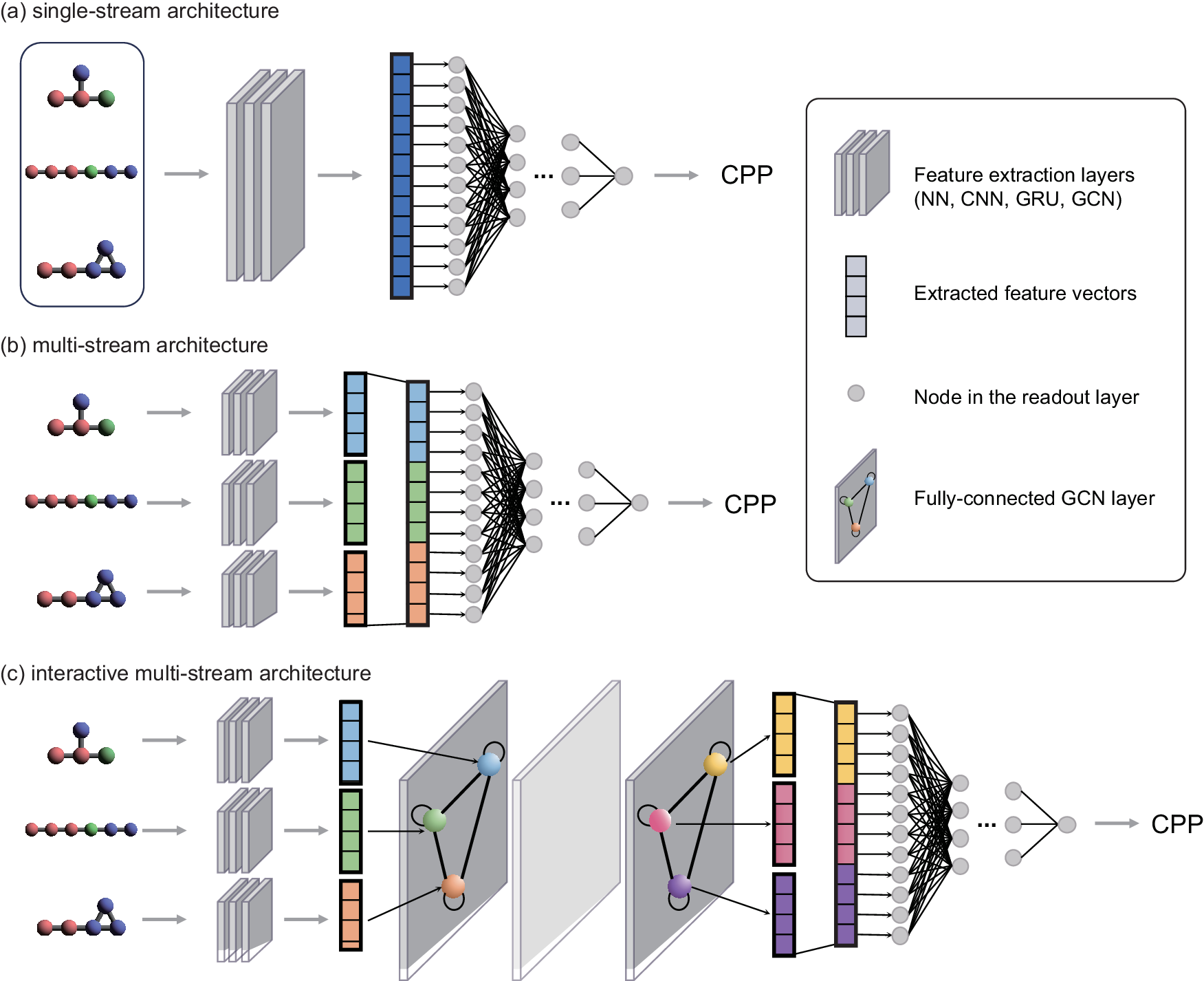}
    \caption{
    Schematic overview of the machine learning architectures employed in this study to predict the critical packing parameter (CPP) of multi-component systems from molecular representations. 
    (a) single-stream architecture: all component features are concatenated into a single input array before feature extraction.
    (b) multi-stream architecture: each component is independently processed, and the extracted features are subsequently integrated for prediction.
    (c) interactive multi-stream architecture: inputs of each component are connected through a fully connected graph convolutional network (GCN) layer to explicitly learn intercomponent interactions.
    The feature extraction layers in these architectures can utilize various machine learning algorithms, including neural networks (NN), convolutional neural networks (CNN), gated recurrent units (GRU), and graph convolutional networks (GCN). CPP denotes the predicted critical packing parameter.
    }
    \label{fig:ml_arch}
\end{figure}

We also investigated three architectural designs for handling multi-component systems, as shown in Fig. \ref{fig:ml_arch}. 
The single-stream (boundary-agnostic) architecture architecture in Fig.~\ref{fig:ml_arch}(a) feeds all components into a single feature-extraction module.  
This design has the lowest computational cost, but once the inputs are combined, the model no longer has a clear way to distinguish which part of the representation comes from which component (except for the GCN case; see below).
For the NN, CNN, and GRU models, the modified-SMILES string of each component is converted into a fixed-length vector \(x\) (with zero-padding as needed), and these vectors are concatenated into a multi-component input \(X\); for example, \(X = [x_1, x_2, x_3]\) for a three-component system.
For the GCN, all molecular graphs are merged into a single disconnected graph by concatenating the adjacency matrices in a block-diagonal manner and stacking the node-feature matrices vertically.
This allows the model to process all components together in a single pass, but information from different components is mixed, which can make it difficult for the model to learn how individual components contribute and interact.

The multi-stream (boundary-aware) architecture in Fig.~\ref{fig:ml_arch}(b) processes each component independently in the feature-extraction stage and then combines the resulting features using a multilayer perceptron (MLP).  
In this design, each component is first mapped to a feature vector \(z\), and these vectors are concatenated into a combined vector \(Z\), which is zero-padded to match the maximum number of components (five in this study).  
For example, \(Z = [z_1, z_2, z_3, 0, 0]\) for a three-component system.  
This architecture keeps the features of each component separate and can flexibly handle different numbers of components by using zero-padding.  
However, the interactions between components are not modeled explicitly; they are learned only indirectly from the concatenated feature vector by the MLP.

The interactive multi-stream (interaction-aware) architecture in Fig.~\ref{fig:ml_arch}(c) extends the multi-stream design by inserting a fully connected GCN between the feature-extraction modules and the final MLP.  
Here, the feature vectors \(z\) for each component serve as node features in a graph whose nodes correspond to components. 
The initial feature matrix is defined as
\[
H^{(0)} =
\begin{bmatrix}
z_1^T \\
z_2^T \\
\vdots
\end{bmatrix},
\]
and updated according to
\[
H^{(l+1)} = \operatorname{ReLU}\!\left(\tilde{D}^{-1/2} \tilde{A}\,\tilde{D}^{-1/2} H^{(l)} W^{(l)}\right),
\]
where \(\tilde{A}\) is the adjacency matrix with self-loops (all components mutually connected),  
\(\tilde{D}\) is the corresponding degree matrix, and \(W^{(l)}\) is the learnable weight matrix at layer \(l\).  
After the graph convolutions, the final node features \(h\) in \(H\) are concatenated, zero-padded, and assembled into a combined vector \(H'\), analogous to \(Z\) in Fig.~\ref{fig:ml_arch}(b); for example, \(H' = [h_1, h_2, h_3, 0, 0]\) for a three-component system.  
This architecture not only preserves the representation of each component, as in the multi-stream design, but also explicitly represents component–component interactions through the fully connected GCN.

\subsection{Hyperparameter tuning}
The machine learning models in this study were built and trained using PyTorch\cite{Paszke2019} and PyTorch Geometric.\cite{Fey2019} The main hyperparameters investigated included the number of convolutional layers in the fully connected GCN (1, 2, 3), the number of fully connected output layers (1, 2, 3), the number of neurons in hidden layers (64, 128, 256) and the learning rates (0.001, 0.01). For feature extraction layers, specific adjustments were made: for NN, the number of layers (1, 2, 3) and neurons in the hidden layers (64, 128, 256); for CNN, the number of filters (32, 64, 128) and kernel sizes (3, 5); for GRU, the number of layers (1, 2) and hidden units (32, 64, 128, 256); and for GCN, the number of convolutional layers (1, 2, 3) and hidden units (32, 64, 128, 256).
The models were trained using the Adam optimizer with a batch size of 128 for 3000 epochs, and the loss function used for evaluation was the mean squared error (MSE) loss. To ensure reliable evaluation during hyperparameter tuning, we used 5-fold cross-validation with a stratified \(k\)-fold procedure. In this approach, the target variable was divided into \(k\) equally spaced bins based on quantiles to create pseudo-class labels, and these labels were used to balance the distribution of the target across all folds. This strategy reduces bias in the target distribution and improves the stability and reliability of model evaluation.
After determining the best architecture through hyperparameter tuning, the final model was retrained on a combined dataset of training and validation data. To evaluate the model’s performance on new data, it was tested using a separate test dataset that had been split in advance. This ensured a robust assessment of the model's ability to generalize.
Full details of the hyperparameter search and all numerical results are provided in the Supporting Information (SI).

\section{Results and discussion}
\subsection{CPP prediction from data with the same number of components}
\begin{figure}[tb]
    \centering
    \includegraphics[width=1.0\linewidth]{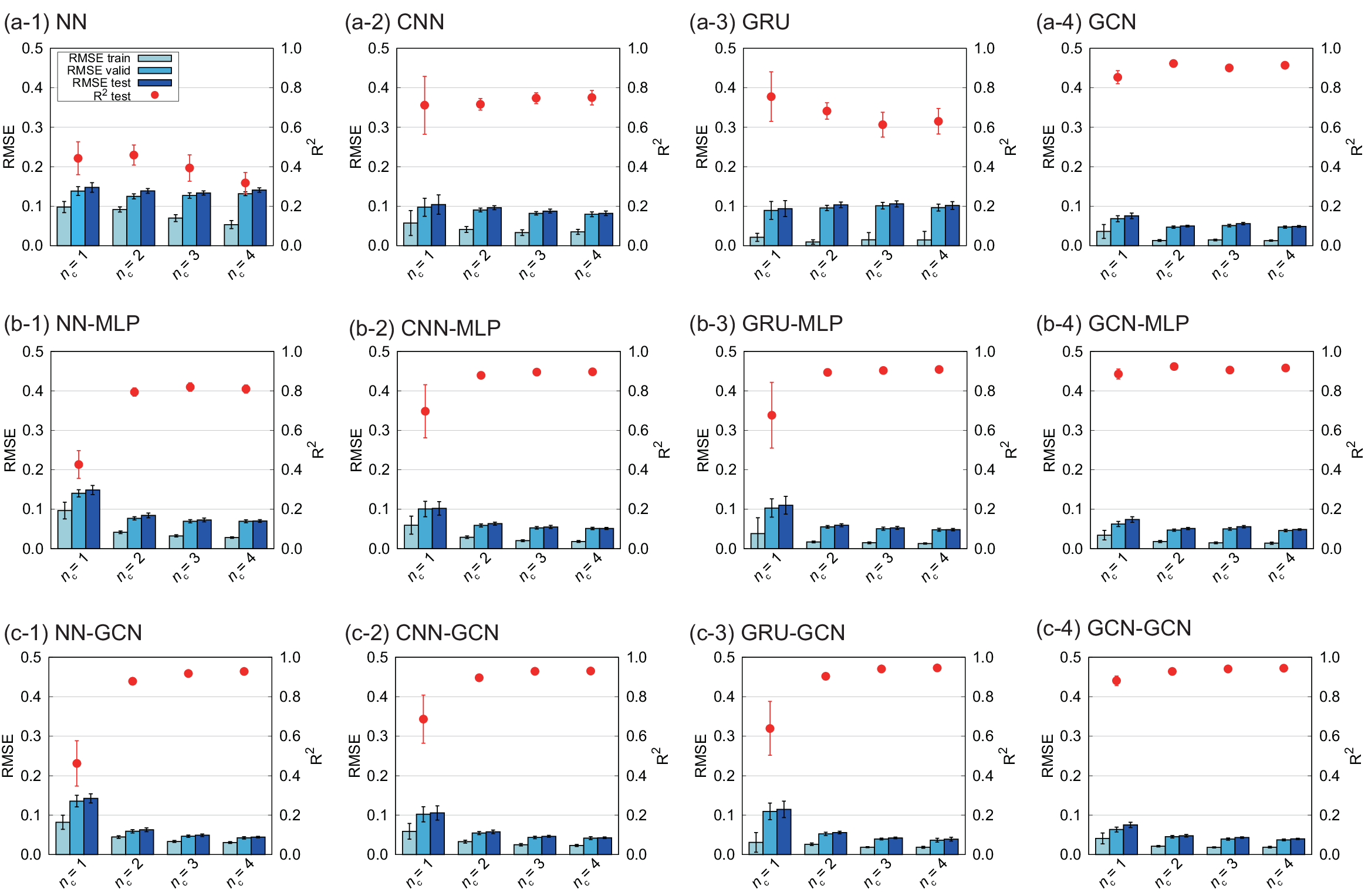}
    \caption{Prediction performance on datasets with the same number of components as used for training ($n_c = 1, 2, 3,$ and $4$).
    Rows (a)–(c) correspond to the architectures shown in Fig.\ref{fig:ml_arch}, and columns (1)–(4) represent the feature-extraction algorithms: (1) NN, (2) CNN, (3) GRU, and (4) GCN. The bar graphs show RMSE values (train/valid/test), and the red circles indicate the $R^2$ scores on the test data. Error bars represent the standard deviation of test performance across 25 trained models.}
    \label{fig:RMSE_same}
\end{figure}

Self-assembly is inherently hierarchical, and accurately predicting multi-component structures therefore requires a machine learning architecture capable of capturing complex multiscale interactions, including both intramolecular and intercomponent interactions.
As a first step toward developing such a model, we evaluated which architectures and algorithms can effectively learn these intricate relationships using the least extrapolative task.
In this setting, models were trained with the systems containing a fixed number of non-water components $n_\text{c}$ and evaluated on the structures with the same $n_\text{c}$.
Twelve models were constructed by combining the three architectures in Fig.~\ref{fig:ml_arch} (single-stream, multi-stream, and interactive multi-stream) with the four feature-extraction algorithms (NN, CNN, GRU, and GCN). 
Dataset with $n_\text{c} = 1, 2, 3, 4$ were split into training, validation, and test sets, and predictions were performed within the same $n_\text{c}$ (Fig.\ref{fig:RMSE_same}).

We first examine the single-stream architecture [Fig.~\ref{fig:ml_arch}(a)], in which all components are concatenated and passed through a single feature-extraction module.  
For single-component systems (\(n_{\mathrm c}=1\)), CNN- and GRU-based models achieve relatively high accuracy, reflecting their ability to learn molecular structures through the local or sequential patterns in the modified-SMILES representation.  
In contrast, NN-based models show low accuracy even for \(n_{\mathrm c}=1\), consistent with our previous finding that simple feed-forward NNs struggle to exploit the sequential order in SMILES-like strings.\cite{Ishiwatari2024}  
For multi-component systems (\(n_{\mathrm c}\ge 2\)), the limitations of the single-stream design become evident.
Concatenating multiple components into a single sequence blurs the boundaries between components, preventing NN-, CNN-, and GRU-based models from effectively learning the intercomponent interactions, resulting in low accuracy [Fig.~\ref{fig:RMSE_same}(a-1)-(a-3)].
Notably, the GCN-based model on the single-stream architecture shows high accuracy for both single- and multi-component systems due to how the molecular structures are represented.
In the graph representation, both single- and multi-component systems appear as a single graph composed of independent molecular subgraphs.
Thus, the message passing occurs only within each subgraph, naturally preventing undesired mixing of information across components.
Consequently, the single-stream GCN model maintains high accuracy for both single- and multi-component systems (\(n_{\mathrm c}=1-4\)), in clear contrast to the NN-, CNN-, and GRU-based single-stream models [Fig.~\ref{fig:RMSE_same}(a-4)].

We next consider the architectures that maintain independent representations for each component, namely the multi-stream and interactive multi-stream architectures [Fig.~\ref{fig:ml_arch}(b,c)].  
In these designs, each component is first encoded independently into a feature vector, and these vectors are then combined to predict the final CPP.
This explicit preservation of component boundaries leads to markedly improved accuracy compared with the single-encoder architecture for multi-component systems (\(n_{\mathrm c} \ge 2\)).
This trend indicates that a hierarchical learning strategy, in which the model first captures intramolecular interactions and then learns intercomponent relationships as self-assembly emerges, is essential for accurate prediction of self-assembled multi-component structures.

Overall, these results reveal two central requirements for CPP prediction in multi-component systems:  
(i) the model must capture molecular structures with sufficient structural detail, and 
(ii) component boundaries must be treated explicitly so that the model can learn how different components interact.
CNN-, GRU-, and GCN-based satisfy the requirement (i) because they can process sequential or graph representations, whereas NN-based models do not.
The single-stream architecture generally fails to meet requirement (ii), except when multi-component systems can be represented as a single graph with disconnected subgraphs.  
In contrast, the multi-stream and interactive multi-stream architectures satisfy (ii) by preserving separate feature representations for each component, and the interactive multi-stream architecture additionally provides an explicit representation of intercomponent interactions.   
Therefore, NN, NN-MLP, and NN-GCN are not considered suitable candidates for generalized CPP prediction in multi-component systems and are excluded from further analysis.  
Based on these insights, we focus in the following sections on seven models: GCN, CNN-MLP, GRU-MLP, GCN-MLP, CNN-GCN, GRU-GCN, and GCN-GCN, which are considered to capture both intramolecular interactions and intercomponent relationships more effectively than the other candidates.

\subsection{Generalization of CPP prediction across different numbers of components}
\begin{figure}[tb]
    \centering 
    \includegraphics[width=0.5\linewidth]{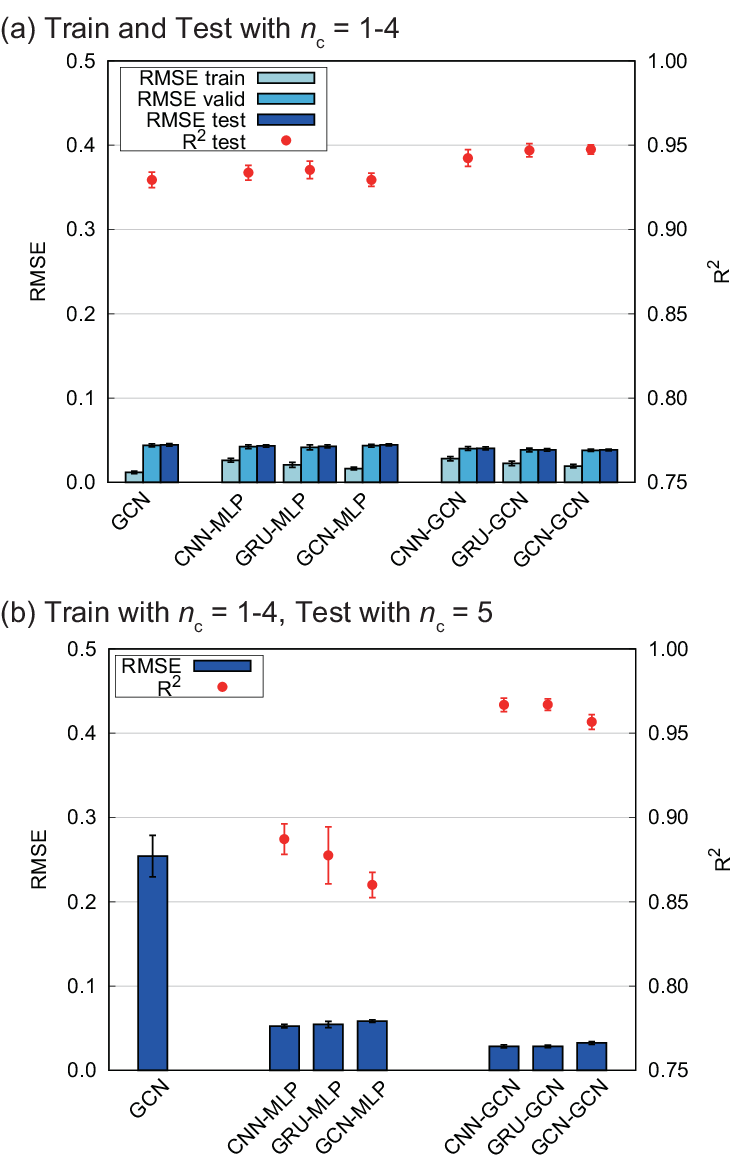}
    \caption{Comparison of seven architectures (GCN, CNN-MLP, GRU-MLP, GCN-MLP, CNN-GCN, GRU-GCN, GCN-GCN). (a) Prediction results when both training and test were conducted using data with $n_c = 1$–$4$ (number of non-water components). (b) Prediction results when training was performed using data with $n_c = 1$–$4$, and test was conducted using data with $n_c = 5$. The $R^2$ value for the GCN was negative and therefore is not shown in this figure. Error bars indicate the standard deviation of test performance across 25 models, generated by five outer splits of the dataset into train/validation and test, and five inner random splits of the training portion into train and validation subsets. }
    \label{fig:RMSE_1234}
\end{figure}

Up to this point, we have shown that the seven models achieve high accuracy when trained and tested on datasets with the same \(n_\text{c}\). 
In practical formulations, however, the number of components is not fixed; a single model must handle mixtures with different \(n_\text{c}\) within one predictive framework. 
As a next step, we therefore examined whether a unified model can make accurate predictions across datasets with diverse component numbers. Concretely, we trained models on datasets containing systems with up to \(n_\text{c} = 4\) and evaluated whether they can simultaneously handle mixtures ranging from \(n_\text{c} = 1\) to \(n_\text{c} = 4\).
To enable such cross-component prediction in architectures with an MLP read-out, the dimensionality of the input to the MLP must be fixed across all samples. 
In our implementation, the fixed-length feature vector \(z\) for each component, obtained from the feature extraction layer, is concatenated into a single vector \(Z\), which is then fed into the read-out MLP. 
To make the length of \(Z\) independent of \(n_\text{c}\), we zero-pad it to the maximum number of components (five in this study), for example \(Z = [z_1, z_2, z_3, 0, 0]\) for a three-component system.
Using this representation, datasets with \(n_\text{c} = 1, 2, 3,\) and \(4\) were merged into a single pool and split into training, validation, and test sets. 
Each model was trained on this mixed-\(n_\text{c}\) dataset and evaluated on a test set that also included mixtures with \(n_\text{c} = 1, 2, 3,\) and \(4\), matching the range of component numbers in the training data. 
All seven models achieved high accuracy (\(R^2 > 0.9\)), comparable to the same-component-number prediction (Fig.~\ref{fig:RMSE_same}), demonstrating that a unified model can treat mixtures with different component counts within one framework and that zero-padding does not noticeably degrade performance [Fig.~\ref{fig:RMSE_1234}(a)]. 

This mixed-\(n_\text{c}\) experiment corresponds to interpolation with respect to the number of components: the model is asked to predict systems whose component counts are already represented in the training data (\(n_\text{c} = 1\text{–}4\)). 
In practical formulations, however, it would be far more valuable if the same model could also perform extrapolation in \(n_\text{c}\), accurately predicting properties of mixtures whose component counts are not present in the training set. 
If such \(n_\text{c}\)-extrapolation were possible, the behavior of higher-order multi-component systems could be inferred from data containing only fewer-component mixtures, greatly reducing the need for exhaustive combinatorial experiments. 
To assess this capability, we next used models trained on datasets with \(n_\text{c} = 1\text{–}4\) to predict the properties of five-component systems (\(n_\text{c} = 5\)), as shown in Fig.~\ref{fig:RMSE_1234}(b).

Among the seven architectures, the standalone GCN model showed a marked decrease in accuracy for \(n_\text{c} = 5\). In this model, all components in a mixture are embedded into a single disconnected molecular graph, and the GCN extracts one global representation for the entire system. This representation appears to depend sensitively on the number of disconnected subgraphs (i.e., components), so changing the component count between training and testing hampers generalization.

In contrast, the three MLP-combined models (CNN–MLP, GRU–MLP, and GCN–MLP) and the three GCN-combined models (CNN–GCN, GRU–GCN, and GCN–GCN) maintained high accuracy in this \(n_\text{c}\)-extrapolation task. 
In all six of these architectures, the same feature extractor is applied independently to each component, and the resulting feature vectors are then combined.
When \(n_\text{c}\) changes, the model simply receives more or fewer of these feature vectors at the aggregation stage, while the way each individual molecular structure is encoded remains the same.
Because the encoding of each component is decoupled from the total number of components, these models are more robust to changes in \(n_\text{c}\) than the standalone GCN. 
Within this group, the three GCN-combined models consistently outperform the MLP-combined models, reflecting the advantage of explicitly modeling intercomponent interactions through the fully connected GCN layer, rather than relying only on an MLP to infer such interactions implicitly from concatenated features.

As an additional assessment of generalization, we also examined extrapolation to systems with fewer components than those used for training. 
In particular, we trained the interactive multi-stream models on multi-component data and tested them on mixtures with a reduced number of components (e.g., \(n_\text{c} = 2,3,4 \rightarrow n_\text{c} = 1\)); in this setting, all three models showed only moderate correlation between predicted and true CPP values, especially for the single-component case (\(n_\text{c} = 1\)), whereas control experiments in which the target \(n_c\) lay within or above the training range (e.g., \(n_\text{c} = 1,3,4 \rightarrow n_\text{c} = 2\) and \(n_\text{c} = 2,3,4 \rightarrow n_\text{c} = 5\)) retained high accuracy (Fig. Sx).  
These results suggest that downward extrapolation to single-component systems is intrinsically more challenging, likely because models trained primarily on multi-component data rely on patterns of intercomponent interactions that are absent at \(n_\text{c} = 1\).  
Detailed parity plots and quantitative metrics for each condition are provided in the Supporting Information (SI).

\subsection{Evaluation of generalization limits in multi-component prediction}
Based on the results presented above, the three architectures that incorporate a fully connected GCN layer (CNN–GCN, GRU–GCN, and GCN–GCN) were able to extrapolate across component counts, accurately predicting CPP for systems with \(n_\text{c} = 5\) even when trained only on data with \(n_\text{c} \le 4\).
However, the previous tests still probed a relatively narrow extrapolation range in terms of component count, corresponding to only a modest step beyond the range of component counts represented in the training data.
A unified prediction framework that genuinely reduces reliance on iterative experimental formulation would need to extrapolate reliably over a broader gap in component count. 
At present, it remains unclear how far this generalization ability extends, in particular to what extent the models can move beyond the component-count range seen during training.
To quantitatively evaluate this capability and identify the limits of generalization achievable with the current multi-component dataset, we conducted additional analyses. In these analyses, we systematically examined how the diversity of component numbers in the training data affects model performance on five-component systems (\(n_\text{c} = 5\)) under two different training configurations.

We designed two training configurations that control the diversity of component numbers in the training data. 
In the first configuration, the total number of training samples is fixed while the range of \(n_\text{c}\) values is widened stepwise. This isolates the effect of \(n_\text{c}\) diversity but inevitably reduces the number of samples for each \(n_\text{c}\), leading to a “sample dilution” effect. 
In the second configuration, the number of samples per \(n_\text{c}\) is held constant. In this case, the total number of training samples increases with the diversity of \(n_\text{c}\), and each component number is sufficiently represented. 
Under both configurations, models were trained on datasets with increasing diversity (\(n_\text{c} = 1\), \(1\text{–}2\), \(1\text{–}3\), and \(1\text{–}4\)) and evaluated on five-component systems. To clarify the role of single-component data, we also trained models on datasets without \(n_\text{c} = 1\) (\(n_\text{c} = 2\), \(2\text{–}3\), and \(2\text{–}4\)) and again tested them on \(n_\text{c} = 5\). 

\begin{figure}[tb]
    \centering 
    \includegraphics[width=1.0\linewidth]{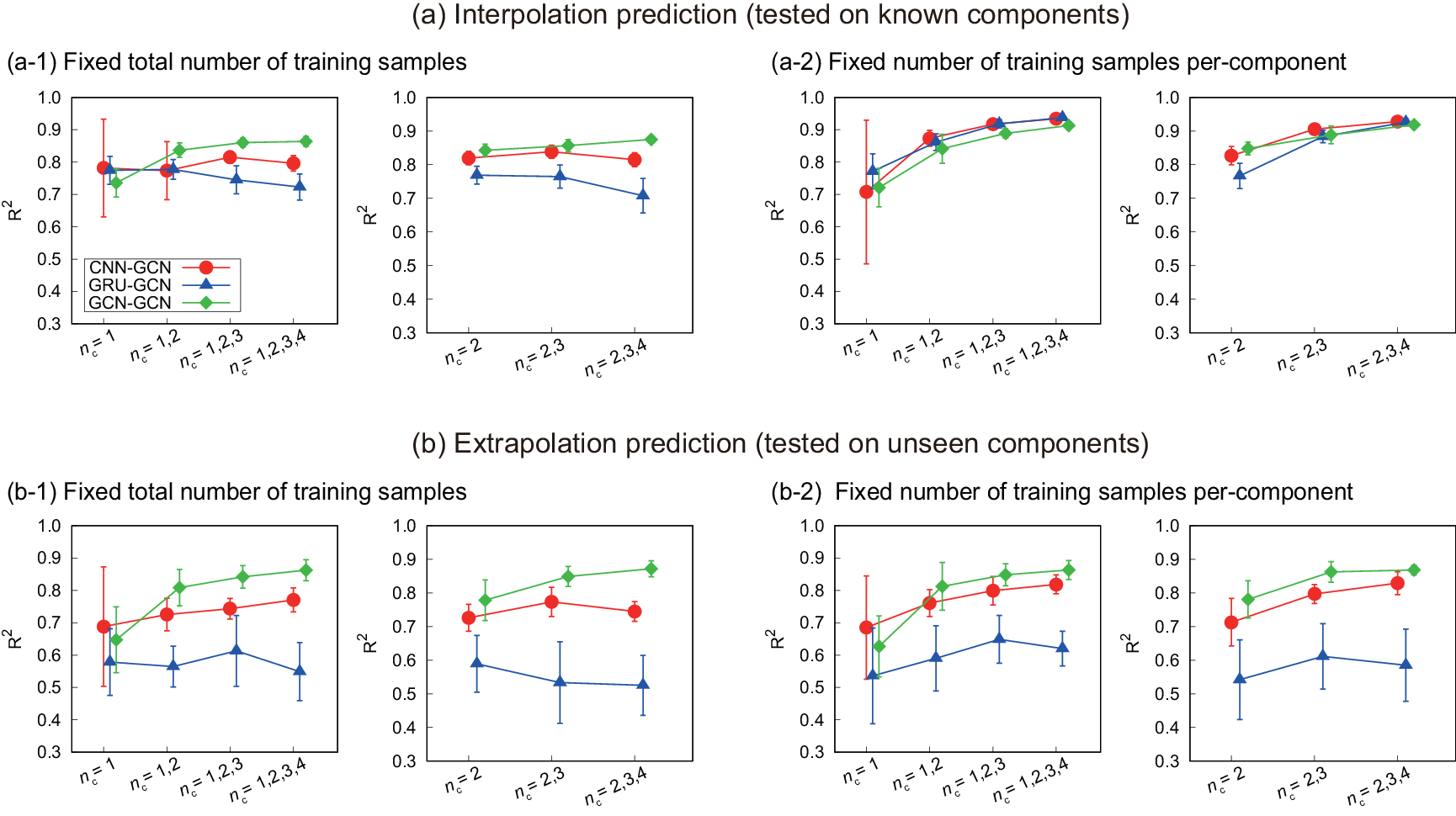}
    \caption{Impact of training-data composition and diversity on the prediction accuracy of CPP values for the 5-component dataset.
    Panels (a) and (b) show interpolation and extrapolation predictions, respectively, where the test data consist of (a) known and (b) unknown components.
    For each prediction type, two training conditions were compared: (a-1, b-1) with a fixed total number of training samples, and (a-2, b-2) with a fixed number of training samples per component (i.e., the total number of training samples increases with the diversity of component counts).
    The horizontal axis represents the range of non-water component numbers ($n_c$) included in the training datasets, while the vertical axis shows the coefficient of determination ($R^2$).
    Error bars indicate the standard deviation across multiple runs.}
    \label{fig:line_plot}
\end{figure}

We analyze the case in which the five-component test systems are composed only of molecules that appear in the training data, that is, prediction for unknown combinations of known components [Fig.~\ref{fig:line_plot}(a)].
For CNN–GCN and GRU–GCN under the fixed-total-sample configuration [Fig.~\ref{fig:line_plot}(a-1)], increasing the diversity of \(n_\text{c}\) yields little or no gain in accuracy. 
Although a broader range of \(n_\text{c}\) should in principle improve generalization, the fixed total sample size makes the data for each \(n_\text{c}\) sparse. 
Under these conditions, the feature vectors produced by the first layer (CNN or GRU) for each component become noisy and unstable, and the fully connected GCN, which operates directly on these features, cannot reliably learn intercomponent interactions. 
When the number of samples per \(n_\text{c}\) is kept constant instead [Fig.~\ref{fig:line_plot}(a-2)], this sample-dilution effect disappears. 
The larger overall dataset stabilizes both the feature representations and the GCN weights, and the prediction accuracy of CNN–GCN and GRU–GCN increases sharply. 
The impact is particularly strong for GRU–GCN, reflecting the fact that recurrent models require ample sequence data to learn robust gating dynamics from the long-range dependencies present in the Modified-SMILES representation.
GCN–GCN behaves differently. 
In this architecture, the first GCN layer directly processes each molecular graph and learns chemically meaningful motifs such as rings and branches. 
As a result, even under the fixed-total-sample configuration [Fig.~\ref{fig:line_plot}(a-1)], the input to the second, fully connected GCN already encodes rich structural information, and the negative impact of sparse data per \(n_\text{c}\) is substantially reduced compared with CNN–GCN and GRU–GCN, whose first-layer features are less explicitly tied to molecular graph structure. 
Increasing the diversity of \(n_\text{c}\) consistently improves performance, because the second GCN layer can observe a wider variety of interaction patterns among components. 
Under the constant-sample-per-\(n_\text{c}\) configuration [Fig.~\ref{fig:line_plot}(a-2)], the additional benefit of increased total data further enhances accuracy, bringing GCN–GCN to the same high level as, or higher than, the other models.

While these analyses quantify generalization over component count and composition for known components, practical applications also require robustness to molecules that never appear in the training set. To assess this stricter form of generalization, we next consider five-component systems composed only of unknown molecular species, corresponding to prediction for unknown combinations of unknown components [Fig.~\ref{fig:line_plot}(b)].
The qualitative trends with respect to training-data diversity are similar to those in Fig.~\ref{fig:line_plot}(a), but the absolute accuracies reveal clear differences among the models. 
CNN–GCN shows a modest drop in \(R^2\) relative to the interpolation setting under both training configurations [Figs.~\ref{fig:line_plot}(b-1) and \ref{fig:line_plot}(b-2)] yet still attains reasonably high accuracy, indicating a moderate ability to transfer learned interactions to unknown components. 
GRU–GCN, in contrast, suffers a more pronounced loss of accuracy: when trained with limited data, its sequence representations do not generalize well beyond the molecules observed in training, and the downstream GCN cannot recover transferable interaction patterns. 
GCN–GCN is the most robust. 
Its accuracy for unknown components in Fig.~\ref{fig:line_plot}(b-1) is almost identical to that in the interpolation case, showing that the graph-based features learned in the first GCN layer transfer effectively to entirely new molecular species.

Overall, these results clarify the generalization limits of the architectures examined in this study. Among the three GCN-combined models, GCN–GCN emerges as the most robust with respect to both the number and the identity of components: it maintains high accuracy when extrapolating from lower to higher component counts and when predicting unknown combinations of unknown components. Notably, even when trained only on two-component systems, GCN–GCN attains \(R^2 \approx 0.8\) for five-component mixtures, indicating that intercomponent interactions learned from a limited set of binary mixtures can be reused to describe higher-order multi-component systems. Within the scope of our simulated dataset, this suggests that accurate multi-component prediction does not necessarily require training data that cover all target component counts or all specific combinations, as long as the model and representation are sufficiently expressive to capture transferable intramolecular and intercomponent relationships.

\section{Conclusions}
In this study, we developed a unified machine-learning framework capable of predicting the self-assembly behavior of multi-component amphiphilic systems directly from the molecular structures of their constituent components. By extending the concept of the critical packing parameter (CPP), originally defined for single-component systems, to multi-component formulations and generating a large, diverse dataset of self-assembled structures through DPD simulations, we systematically evaluated twelve combinations of feature-extraction algorithms and model architectures. Our results demonstrated that models incorporating a fully connected GCN layer can explicitly learn intercomponent interactions and consistently outperform other architectures across all component numbers.

Same-component-number predictions revealed that accurately distinguishing component boundaries and capturing intramolecular structural features are both essential for high-accuracy CPP prediction in multi-component systems. In particular, the GCN–GCN model achieved consistently high accuracy for all systems with 
$n_c=1 – 4$, effectively learning both intramolecular and intercomponent relationships through graph-based molecular representations.
In addition, models trained on datasets that combined systems with different \(n_\text{c}\) values maintained similarly high accuracy across all component numbers, showing that mixtures with varying numbers of components can be handled within a single predictive framework; in this setting, the zero-padding scheme used to standardize input length did not noticeably degrade performance.

A key result of this work is the strong extrapolative capability of the GCN–GCN model, which accurately predicted CPP values for systems with component numbers and molecular species not present during training. Models trained only on data with $n_c=1 – 4$ successfully predicted CPP values for previously unknown five-component systems. Moreover, even when trained without single-component data, the model maintained strong predictive performance for multi-component mixtures. Most notably, the GCN–GCN model demonstrated excellent extrapolation to five-component systems composed entirely of molecular species absent from the training set, underscoring its potential as a composition-agnostic, molecular-level predictive framework.

We further showed that high predictive accuracy for complex multi-component systems can be achieved using only binary-mixture training data. This result is particularly significant given the substantial experimental cost of collecting data for higher-order mixtures: once pairwise interactions are sufficiently learned, our model can reliably infer the behavior of systems containing three or more components. Combined with the fact that the model requires only molecular structures as input, these results highlight a practical and sustainable strategy for materials development, enabling rapid virtual screening and reducing reliance on extensive trial-and-error experimentation.

Overall, this work presents the first unified machine-learning framework capable of predicting self-assembly in multi-component amphiphilic systems with arbitrary component numbers, including systems composed of previously unknown molecular species. We anticipate that this approach will open new opportunities for the inverse design of multi-component soft materials, including drug-delivery carriers, surfactant formulations, and polymer–colloid hybrid systems. Future extensions incorporating concentration, temperature, and solvent conditions may further broaden the applicability of this framework and support the rational design of increasingly complex self-assembling materials.


\begin{suppinfo}
Additional details regarding the simulation dataset, including the distribution of CPP values, hyperparameter optimization results, training loss curves, and supplementary evaluations of generalization across different numbers of components, are provided in the Supporting Information.
\end{suppinfo}

\bibliography{ref}

\end{document}